\documentclass[12pt, a4paper, twoside, final]{article}

\usepackage{natbib}	

\usepackage{amsmath}

\usepackage{geometry}

\usepackage{amsthm}

\usepackage{amsfonts}

\usepackage{bbm}

\DeclareMathOperator{\disc}{disc}

\DeclareMathOperator{\herdisc}{herdisc}

\DeclareMathOperator{\lindisc}{lindisc}

\DeclareMathOperator{\wdisc}{wdisc}

\DeclareMathOperator{\herwdisc}{herwdisc}

\newcommand{\abs}[1]{\left\lvert #1 \right\rvert}

\newcommand{\abse}[1]{\bigg| #1 \bigg|}

\newcommand{\floors}[1]{\left\lfloor #1 \right\rfloor}

\newcommand{\clength}[1]{\abs{#1}_c}

\newcommand{\HH}{{\mathcal{H}}}

\newcommand{\EE}{{\mathcal{E}}}

\newcommand{\N}{\mathbb{N}}

\newtheorem{theorem}{Theorem}[section]

\newtheorem{lemma}[theorem]{Lemma}

\newtheorem{fact}[theorem]{Fact}

\newenvironment{proof-idea}{\noindent{\textit{Proof Idea.}}\hspace*{1em}}

\theoremstyle{definition}

\renewcommand{\baselinestretch}{1.4}

\begin{document}

\title{Hereditary Discrepancies in Different Numbers of Colors II}

\author{Benjamin Doerr\thanks{Max--Planck--Institut f\"ur Informatik,
  Saarbr\"ucken, Germany.} \and Mahmoud Fouz\thanks{Universit\"at des Saarlandes, Saarbr\"ucken, Germany}}

\maketitle

\begin{abstract} 

  We bound the hereditary discrepancy of a hypergraph $\HH$ in two colors in terms of its hereditary discrepancy in $c$ colors. We show that $\herdisc(\HH,2) \le K c \herdisc(\HH,c)$, where $K$ is some absolute constant. This bound is sharp.

\end{abstract}

\section{Introduction}

Discrepancy theory examines uniformity problems of combinatorial and geometric objects as well as sequences of numbers.

The discrepancy problem for hypergraphs  is to color the vertices of a hypergraph with  fixed number of colors such that each hyperedge contains about the same number of vertices in each color. This problem is the subject of  a number of famous results discovered in the last four decades, among them the theorem of Ghouila-Houri \cite{ghou}, the Beck-Fiala theorem \cite{bf}, and Spencer's `six standard deviations' bound \cite{spencer}.

Surprisingly, almost all research so far regarded hypergraph discrepancies for two colors only. It was only recently that the general case was formally introduced and investigated by Srivastav and the first author~\cite{wirmcol}. Their results revealed a remarkable peculiarity. Although many of the classical results could be generalized to the multi-color case, it turned out that there is no general relationship between the discrepancies of a hypergraph in different numbers of colors. In particular, it was shown in \cite{doerr-disc} that there exist families of hypergraphs with zero discrepancy in certain numbers of colors and high discrepancy in all others.

This dichotomy could be solved 

in~\cite{doerr-hereditary}. There it was shown that the \emph{hereditary discrepancy}, which is the maximum discrepancy among the induced subhypergraphs, is almost independent of the number of colors. In this paper, we further sharpen this relation to bounds that are tight up to constant factors (independent of the numbers of colors and the size of the hypergraph).

\section{Combinatorial Discrepancy Theory}

In this section, we give a short introduction to  discrepancies of hypergraphs and matrices. The interested reader is invited to consult the nice survey on discrepancy theory by Beck and S\'{o}s \cite{sos} for further information.

\subsection{Combinatorial Discrepancies of Hypergraphs}

A (finite) \emph{hypergraph} $\mathcal{H}$ is an ordered pair $(V, \mathcal{E})$ where $V$ is a (finite) set of \emph{vertices} and $\mathcal{E} \subseteq 2^V$ is a set of \emph{(hyper)edges}. In the following, we restrict ourselves to finite hypergraphs without explicit notice. We also stick to the conventions to denote the number of vertices by $n$ and the number of edges by $m$.

Let $c \geq 2$. A \emph{$c$--coloring} of $\HH$ is simply a mapping $\chi : V \to M$, where $M$ is some set of cardinality $c$. Typically, we have $M = [c] := \{1, \ldots, c\}$. The hypergraph $\mathcal{H}$ is ideally colored by $\chi$ if each color $d$ occurs equally often in each hyperedge, that is, if $\abs{\chi^{-1}(d)\cap E} = \frac{1}{c}\abs{E}$ for all  $d\in [c]$ and $E\in \mathcal{E}$.  Hence, we define  the \emph{$c$--color discrepancy of $\mathcal{H}$ with respect to $\chi$} by

\begin{equation*}
 \disc(\mathcal{H},\chi, c) := \max_{d\in [c], E\in \mathcal{E}}\abs{|\chi^{-1}(d)\cap E| - \tfrac{1}{c}\abs{E}}.
\end{equation*}

The \emph{$c$--color discrepancy of $\mathcal{H}$} is now given by

\begin{equation*}
 \disc(\mathcal{H},c) := \min_{\chi:V\to[c]}\disc(\mathcal{H},\chi,c).
\end{equation*}

To get some intuition for this concept, let us regard the following hypergraph $\HH_n$. Let $A, B$ be disjoint sets of cardinality $n$ and $\HH = (A \cup B, \EE)$ with $\EE = \{E \subset A \cup B \mid |E \cap A| = |E \cap B|\}$. 

\begin{fact}

For all $n \in \N$, the discrepancy of $\HH_n$ satisfies the following.

\begin{enumerate}

	\item[(a)] $\disc(\HH_n,2) = 0$.

	\item[(b)] For all $c > 2$, $\disc(\HH_n,c) \ge \frac n c$.

\end{enumerate}

\end{fact}

In particular, we see that a hypergraph may have very different discrepancies in different numbers of colors. As we will see, this is not possible for the hereditary discrepancy.

Let $V_0 \subseteq V$ be a subset of the vertices. Then $\HH_{\rvert V_0} := (V_0, \{E \cap V_0 \mid E \in \EE\})$ is called \emph{induced subhypergraph} of $\HH$. The \emph{$c$--color hereditary discrepancy} of $\HH$ is the maximum $c$--color discrepancy among the induced subhypergraphs: 

\begin{equation*}
 \herdisc(\HH,c) = \max_{V_0\subseteq V} \disc(\HH_{\rvert V_0},c). 
\end{equation*}

Again, we roughly estimate the hereditary discrepancy of $\HH_n$ in different number of colors.

\begin{fact}

  For all $n \in \N$, \[\tfrac nc \le \herdisc(\HH,c) \le \tfrac{2n}c.\]

\end{fact} 

That the hereditary discrepancies are similar in different numbers of colors, was proven by the first author~\cite{doerr-hereditary}.

\begin{theorem}\label{talt}

  For all hypergraphs $\HH$ and all numbers $a,b \in \mathbb{N}_{\ge 2}$ of colors, 

  \[\herdisc(\HH,b) \le a^2 (b-1) \herdisc(\HH,a).\]

\end{theorem}

In Section~\ref{secmain}, we sharpen this bound to $\herdisc(\HH,b) \le K a \herdisc(\HH,a)$ for an absolute constant $K$. The linear dependence on $a$ is necessary, as shown in Section~\ref{seclower}.

\subsection{Discrepancies of Matrices}

We can describe a hypergraph via its \emph{incidence matrix}. Let $v_1, \dots, v_n \in V$ be the vertices and $E_1, \dots, E_m \in \mathcal{E}$ be the edges of a hypergraph $\mathcal{H}$. Then the incidence matrix with respect to these orderings of $V$ and $\mathcal{E}$ is the $m\times n$ matrix $A = (a_{ij})$ with

\begin{displaymath}
 a_{ij} = \begin{cases}

           1 & \text{if $v_j \in E_i$ }, \\

	   0 & \text{else}.

          \end{cases}
\end{displaymath}

For a given $c$--coloring $\chi : V \to [c]$ of a hypergraph, we define an associated $c$--coloring of its incidence matrix as a mapping $p: [n] \to [c]$ with $p(i) := \chi(v_i)$ for all $i\in [n]$. The following definitions extend the discrepancy notions to matrices in the sense that the discrepancies of a hypergraph and its incidence matrix are the same. For arbitrary matrix  $A\in \mathbb{R}^{m\times n}$ and $c$--coloring $p:[n]\to[c]$, we define

\begin{align*}
 \disc(A,p,c) &:= \max_{d\in [c], i\in [m]} \abs{\sum_{j\in p^{-1}(d)}a_{ij}-\frac{1}{c}\sum_{j\in[n]}a_{ij}},\\
 \disc(A,c) &:= \min_{p:[n]\to [c]}\disc(A,p,c),\\
 \herdisc(A,c) &:= \max_{A_0 \subseteq A} \disc(A_0,c).
\end{align*}

Here we write $A_0 \subseteq A$ to denote that $A_0$ is a submatrix of $A$. In the following, we will mainly work with the more general concept of matrix discrepancies.

\subsection{Weighted Discrepancies}

The discrepancy notions introduced above all refer to the problem of partitioning into equal sized parts. Occasionally, and in particular in this paper, a more general concept is helpful. In the language of hypergraphs, the notion of weighted discrepancy refers to the problem of coloring the vertices in such a way that each hyperedge has a certain fraction of its vertices in each color class.

Fortunately, we need such weighted discrepancies only for two colors. Here the problem is slightly simpler as we only have to regard one color. Note that an excess of vertices in one color in some hyperedge yields the same number of vertices missing in the other color.

In the matrix language, we define the following. A mapping $p : [n] \to [0,1]$ shall be called \emph{floating coloring}. For two floating colorings $p$ and $q$, we define their discrepancy by \[d_A(p,q) := \|A(p,q)\|_\infty = \max_{i \in [m]} \bigg|\sum_{j \in [n]} a_{ij} (p(j) - q(j))\bigg|.\]

Let $z \in [0,1]$. Denote by $\bf{1}_n$ the unique mapping $[n] \to \{1\}$. Then

 \begin{align*}
 \wdisc(A,z) &:= \min_{q:[n]\to\{0,1\}} d_A(z {\mathbf 1}_n,q),\\
 \wdisc(A,2) &:= \max_{z\in [0,1]}\wdisc(A,z),\\
 \herwdisc(A,2) &:= \max_{A_0\subseteq A}\wdisc(A_0,2).
\end{align*}

Having introduced the (hereditary) weighted discrepancy, we recognize the `usual' discrepancy notions as a special case.

\begin{fact}

\label{other discrepancy definition}

 For all matrices $A$, we have 

\begin{align*}
 \disc(A,2) &=\wdisc(A,\tfrac{1}{2}),\\
 \herdisc(A,2) &= \herwdisc(A,\tfrac{1}{2}).
\end{align*}

\end{fact}

Surprisingly, the weighted discrepancy problem is not much harder than the aforementioned special case. The following theorem follows from a famous result by Beck and Spencer \cite{BS} as well as Lov\'{a}sz, Spencer, and Vesztergombi \cite{LSV}

\begin{theorem}
\label{weighted vs. normal}
For all matrices $A$, we have
\begin{equation*}
\herwdisc(A,2) \leq 2\herdisc(A,2).
\end{equation*}
\end{theorem}

The additional strength of the notion of weighted discrepancies is visible e.g. in the following result, which we will later also need.

\begin{theorem}[Doerr, Srivastav \cite{wirmcol}]

\label{t2col}

 For an arbitrary matrix $A \in \mathbb{R}^{m \times n}$ and any $c \in \mathbb{N}_{\geq 2}$,

\begin{equation*}
 \herdisc(A,c) \leq 2.0005 \; \herwdisc(A,2).
\end{equation*}

\end{theorem}

\section{Discrepancies in Different Numbers of Colors}\label{secmain}

In this section, we improve Theorem~\ref{talt} by proving the following bound.

\begin{theorem}

\label{main theorem}

For any matrix $A \in \mathbb{R}^{m \times n}$  and arbitrary $ a,b \in \mathbb{N}_{\geq 2}$, we have 

	\begin{equation*} 
		\emph{herdisc}(A,b)  \leq  K a \; \emph{herdisc}(A,a),
	\end{equation*}

	where $K$ is some absolute constant less than $2.0005$.

\end{theorem}

For the proof, we will take a detour through two colors. Since by Theorem~\ref{t2col}, the hereditary discrepancy in $b$ colors is at most of the order of the hereditary weighted discrepancy in two colors, it remains to bound the hereditary weighted discrepancy in two colors by the $a$--color hereditary discrepancy. Hence, the main part of this paper is to prove the following theorem.

\begin{theorem}

\label{main lemma}

For any matrix $A \in \mathbb{R}^{m \times n}$  and arbitrary $c \in \mathbb{N}_{\geq 2}$, we have 

	\begin{equation*} 
		\herwdisc(A,2)  \leq  c\: \herdisc(A,c).
	\end{equation*}

\end{theorem}

The proof of Theorem \ref{main theorem} now is an easy corollary.

\begin{proof}[Proof of Theorem \ref{main theorem}]

We first bound the hereditary discrepancy in $b$ colors by the 2--color weighted hereditary discrepancy and afterwards bound the latter by the hereditary discrepancy in $a$ colors. Hence, we get 

 \begin{align*}
  \herdisc(A,b) & \leq  2.0005\; \herwdisc(A,2) &&\text{by Theorem \ref{t2col}} \\
				& \leq 2.0005a\; \herdisc(A,a) && \text{by Theorem \ref{main lemma}}.
 \end{align*}

\end{proof}

\subsection{A Simple Case}

Both to give some more insight and to use it later, let us first analyze the case where $a$ is a multiple of $b$.

\begin{lemma}

\label{simple case lemma}

 Let $a = k\cdot b$ for some $k\in \mathbb{N}$. Then

\begin{equation*}
 \herdisc(A,b)  \leq  k \; \herdisc(A,a).
\end{equation*}

\end{lemma}

\begin{proof}

 Let $A_0 \in \mathbb{R}^{m_0 \times n_0}$ be an arbitrary submatrix of $A$. Choose an $a$--coloring $p:[n_0]\to[a]$ with $\disc(A_0, p, a) \leq \herdisc(A,a)$. We will define a $b$--coloring $q:[n_0]\to[b]$ by clubbing together $k$ color classes of $p$ respectively, i.e., 

\begin{equation*}
 q(i) \equiv p(i) \mod b.
\end{equation*}

We obtain $\disc(A_0, b)\leq \disc(A_0, q,b) \leq k\; \herdisc(A,a)$. Since we chose $A_0$ arbitrarily, the claim follows. 

\end{proof}

If $a$ is not a multiple of $b$, we cannot evenly combine color classes as done above. We could apply Lemma \ref{simple case lemma} only using the $b\lfloor\frac{a}{b} \rfloor$ largest color classes and recursively repeat this procedure on the vertices that were not colored so far.

 However this recursive approach might take $\log_{1/(1-\frac{b}{a}\lfloor \frac{a}{b}\rfloor)}(n_0)\leq \log_2(n_0)$ iterations, yielding a bound with logarithmic dependence on the number of vertices.

\subsection{The Main Result}

In this section, we bound the 2--color weighted hereditary discrepancy in terms of the hereditary discrepancy in $c$ colors, that is, we prove Theorem \ref{main lemma}.

For $c$ even, this can be deduced from Theorem \ref{weighted vs. normal} and Lemma \ref{simple case lemma} since 
\begin{align*}
\herwdisc(A,2) &\leq 2\herdisc(A,2) &&\text{by Theorem \ref{weighted vs. normal}}\\ 
	&\leq c\herdisc(A,c) && \text{by Lemma \ref{simple case lemma}}.
\end{align*}
Hence, the crucial case is that $c$ is odd. We show that any constant floating coloring $p := z {\mathbf 1}_n$, where $z$ has finite $c$--ary expansion of length $\ell$, can be `rounded' to a floating coloring $q$ having $c$--ary length at most $\ell -1$ and $d_A(p,q) = O(c^{-\ell+2})$. Removing the obstacle that $q$ is not necessarily a constant floating coloring, we can iterate this rounding procedure until we obtain a true coloring.

The following lemma describes one iteration of the rounding process. If some number $x$ can be written as $x = \sum_{i = 0}^\ell a_i c^{-i}$ with $\ell \in \N$ and $a_0, \ldots, a_\ell \in \{0, 1, \ldots, c-1\}$, we say that $x$ has $c$--ary expansion of length $\ell$. By $\clength{x}$ we denote the length of a shortest $c$--ary expansion of $x$. We set $\clength{x} = \infty$, if $x$ has no finite $c$--ary expansion. For $\ell \in \N_0$ and $N \subseteq \N$, we define 

\begin{eqnarray*}
  M_{c,\ell} &:=& \left\{ x \in \left[0,1\right] \vert \clength{x} \leq \ell\right\},\\
  C^{N}_{c,\ell} &:=& \left\{p  \,\vert\, p: N\to M_{c,\ell}\right\}.	
\end{eqnarray*}

\begin{lemma}

\label{rounding in principle} 

 Let $c$ be odd, $t\in M_{c,\ell}$ and $p \in C^{[n]}_{c,\ell}$ with $p(i) = t$ for all $i\in [n]$. Then there exists a $p' \in C^{[n]}_{c,\ell-1}$ such that

  \begin{equation*}
\emph{$d_A$} \left(p,p'\right)   \leq  \tfrac{1}{2}(c-1) c^{-\ell+1}  \herdisc(A,c). 
\end{equation*} 

\end{lemma}

\begin{proof}

Let $t = \sum_{k=0}^{\ell}t_{k} c^{-k}$ for some $t_k \in \left\{0,\dots,c-1\right\}$ denote the $c$-ary expansion of $t$ \footnote{Remember that this expansion of $t$ is unique.} and ${\lfloor t\rfloor}_{\ell-1}:=\sum_{k=0}^{\ell-1}{t_{k} c^{-k}} \in M_{c,\ell-1}$ be its rounding of $c$--ary length $\ell-1$.

Choose a $c$--coloring $q : [n] \to [c]$ such that disc$\left(A, q ,c\right) \leq \mbox{herdisc}\left(A , c \right)$. Denote by $J_i := \left\{j\in [n] \,\vert\, q\left(j\right) = i\right\}$ for $i \in [c]$ the partition classes defined by $q$. Let $J_{[t_{\ell}]}:= \bigcup_{k=1}^{t_{\ell}}{J_k}$ be the union of the first $t_{\ell}$ partition classes. For all $j \in [n]$ put

\begin{equation*}
 p'(j) := \begin{cases}

{\lfloor t\rfloor}_{\ell-1}  + c^{-\ell+1} & \text{if $j \in J_{[t_{\ell}]}$}, \\
{\lfloor t\rfloor}_{\ell-1} & \text{else}.

\end{cases}
\end{equation*}

Note that $p'(j) \in M_{c,\ell-1}$ for all $j \in [n]$. Hence $p' \in C^{[n]}_{c,\ell-1}$.

For all $i \in [m]$, we compute

\begin{align*}
	 \lefteqn{\abse{\sum_{j\in[n]} a_{ij} \left(p(j) - p'(j)\right)}} \\ 
	&=  \abse{\sum_{j\in J_{[t_{\ell}]}} a_{ij}\left(p(j) - {{\lfloor t }\rfloor}_{\ell-1} - c^{-\ell+1}\right) + \sum_{j\in [n] \setminus J_{[t_{\ell}]}} a_{ij}\left(p(j) - {\lfloor t\rfloor}_{\ell-1}\right)} \\
	&=  \abse{\sum_{j\in J_{[t_{\ell}]}}  a_{ij}\left(c^{-\ell}t_{\ell} - c^{-\ell+1}\right) + \sum_{j\in [n] \setminus J_{[t_{\ell}]}} a_{ij} c^{-\ell}t_{\ell}}\\ 
	&=  c^{-\ell+1}\abse{\sum_{j\in J_{[t_{\ell}]}} a_{ij}\left(c^{-1}{t}_{\ell} - 1\right) + \sum_{j\in [n] \setminus J_{[t_{\ell}]}} a_{ij} c^{-1}t_{\ell}} \\
	&=  c^{-\ell+1}\abse{\sum_{j\in [n]} a_{ij} c^{-1}t_{\ell} - \sum_{j\in J_{[t_{\ell}]}} a_{ij}} \\
	&=  c^{-\ell+1} \abse{\sum_{e \in [t_{\ell}]}\bigg({\sum_{j\in [n]} a_{ij} c^{-1} - \sum_{j\in J_{e}} a_{ij}}\bigg)} \\
	&\tag{1}\leq  c^{-\ell+1} \min\{t_{\ell},c-t_{\ell}\} \herdisc(A,c)\\
	&\tag{2}\leq  \tfrac{1}{2}(c-1) c^{-\ell+1} \herdisc(A,c)
\end{align*}

Here (1) follows from the fact that 

\begin{equation*}
\bigg| \sum_{e \in [t_{\ell}]}\bigg({\sum_{j\in [n]} a_{ij} c^{-1} - \sum_{j\in J_{e}} a_{ij}}\bigg)\bigg| = \bigg| \sum_{e \in [c]\setminus[t_{\ell}]}\bigg({\sum_{j\in [n]} a_{ij} c^{-1} - \sum_{j\in J_{e}} a_{ij}}\bigg)\bigg|.
\end{equation*}

For (2), we note that $c$ is odd.  

\end{proof}

While we assumed all values of $p$ to be equal, its rounding $p'$ may take two different values. To apply Lemma \ref{rounding in principle} iteratively, we generalize our result for that purpose.

\begin{lemma}

\label{one round in general}

 Let $c$ be odd and $p \in C^{[n]}_{c,\ell}$ taking $\tau$ different values. Then there exists a $q \in C^{[n]}_{c,\ell-1}$ such that

\begin{equation*}
\emph{$d_A$} \left(p,q\right) \leq   \tfrac{1}{2}(c-1) c^{-\ell+1} \tau \emph{herdisc}\left(A, c\right) 
\end{equation*}

\end{lemma}

\begin{proof}

We apply Lemma \ref{rounding in principle} on subinstances of the rounding problem where $p$ is constant. Let $t^{(1)}, \dots, t^{(\tau)} \in M_{c,\ell}$ be an enumeration of the different values of $p$. For all $k\in [\tau]$, define $\mathcal{J}_k := \left\lbrace j\in [n] \,\vert\, p(j) = t^{(k)}\right\rbrace$. Writing $A_j$ for the $j$-th column of $A$, we denote by $A_{\rvert \mathcal{J}_k}$ the submatrix consisting of the columns $A_{j}$ of $A$ with $j \in \mathcal{J}_k$. \footnote{We adopt the formal view that a matrix $A\in\mathbb{R}^{m\times n}$  is a function $A: [m]\times[n]\to \mathbb{R}$. Hence, the submatrix $A_{\vert \mathcal{J}}$ is simply the restriction $A_{\vert \mathcal{J}}:[m]\times \mathcal{J}; (i,j)  \mapsto A(i,j)$ for all $i\in[m]$ and $j\in \mathcal{J}$.}

Now $p_{\rvert \mathcal{J}_k}$ is a constant floating coloring for $A_{\rvert \mathcal{J}_k}$. Apply Lemma \ref{rounding in principle} on $A_{\rvert \mathcal{J}_k}$ and $p_{\rvert \mathcal{J}_k}$ to obtain a floating coloring $q^{\left(k\right)} \in C_{c,\ell-1}^{\mathcal{J}_k}$ with \begin{align*}
d_{A_{\vert \mathcal{J}_k}} (p_{\vert \mathcal{J}_k},q^{\left(k\right)})  & \leq  \tfrac{1}{2}(c-1) c^{-\ell+1} \herdisc(A_{\vert \mathcal{J}_k}, c)\\
& \leq  \tfrac{1}{2}(c-1) c^{-\ell+1} \herdisc\left(A, c\right).
\end{align*}

Now let $q$ be the union of the $q^{(k)}$, that is,  $q\in C^{[n]}_{c,\ell-1}$ such that  $q(j) = q^{(k)}(j)$ for all $j\in \mathcal{J}_k$ and $k\in [\tau]$.

Since the $\mathcal{J}_k$ form a partition of $[n]$, we have

\begin{align*}
d_{A}\left(p,q\right) & =  \max_{i\in [m]} \bigg| \sum_{k\in[\tau]} \sum_{j\in J_k} a_{ij} \left(p(j)-q(j)\right)\bigg| \\
& \leq  \sum_{k\in[\tau]}  \max_{i\in [m]} \bigg| \sum_{j\in[\mathcal{J}_k]} a_{ij} \left(p(j)-q(j)\right)\bigg| \\
& \leq  \sum_{k\in[\tau]} d_{A_{\vert \mathcal{J}_k}} (p_{\vert \mathcal{J}_k},q^{\left(k\right)} )\\
& \leq  \tfrac{1}{2}(c-1) c^{-\ell+1} \tau  \herdisc(A, c).
\end{align*}

\end{proof}

Starting with a floating coloring $p \in C^{[n]}_{c,\ell}$, we iteratively apply the lemma above $\ell$ times to get a true coloring $q$. The crucial observation now is that if  our initial floating coloring $p$ is constant, then all colorings constructed by applying Lemma~\ref{one round in general} take at most two different values. In other words, we always have $\tau \le 2$ when invoking Lemma~\ref{one round in general}.

\begin{lemma}

\label{number of different values is two}

Let $c$ be odd, $t \in M_{c,\ell}$ and $p = t {\mathbf 1}_n$. Denote by $q^{(i)}$ the vector obtained after the $i$-th iteration of applying Lemma \ref{one round in general} on $p$. Then the number of different values of $q^{(i)}$ is at most 2 for all $i \in \{0, \dots, \ell\}$. 

\end{lemma}

\begin{proof}

We distinguish two states of the rounding algorithm. In state 1, there are at most two different values of the current vector $q^{(i)}$ and all values differ only in the last digit, that is, $q^{(i)}$ takes at most two values $t^{(1)}$ and $t^{(2)}$ such that $\floors{t^{(1)}}_{\ell-i-1} = \floors{t^{(2)}}_{\ell-i-1}$. In state 2, the current vector $q^{(i)}$ takes two values $t^{(1)}$ and $t^{(2)}$ of different $c$-ary length such that $t^{(1)} = t^{(2)} + c^{-l+i}$, $\abs{t^{(1)}}_{c} = \ell-i-o$ and $\abs{t^{(2)}}_{c} = \ell-i$ for some $o\in [l-i]$. Furthermore, if $t^{(2)} = \sum_{k=0}^{l-i} t^{(2)}_{k} c^{-k}$ denotes the $c$-ary expansion of $t^{(2)}$, then $t^{(2)}_{\ell-i-k} = c-1$ for all $k\in \{0,\dots, o-1\}$ and $\floors{t^{(1)}}_{\ell-i-o-1} = \floors{t^{(2)}}_{\ell-i-o-1}$. We show that the algorithm is always in one of these two states and in particular that no other state is reached.

Clearly, the algorithm starts in state 1. 

Assume that after iteration $i$ the algorithm is in state 1 and let $t^{(1)}$ and $t^{(2)}$ be defined as above for state 1, where $t^{(1)} = \sum_{k=0}^{\ell-i}t^{(1)}_{k}c^{-k}$ denotes the $c$-ary expansion of $t^{(1)}$. By applying the procudure described in Lemma \ref{one round in general}, we get $q^{(i+1)}(j) = \floors{t^{(1)}}_{\ell-i-1} + c^{-\ell+i+1}$ or $q^{(i+1)}(j) = \floors{t^{(1)}}_{\ell-i-1}$ for all $j \in [n]$. Note that we have exploited the fact that $\floors{t^{(1)}}_{\ell-i-1} = \floors{t^{(2)}}_{\ell-i-1}$. We now distinguish two cases.\\ 
If $\floors{t^{(1)}}_{\ell-i-1} + c^{-\ell+i+1}$ results in no carry over, that is, if $t^{(1)}_{l-i-1} \neq c-1$, then we have $\floors{q^{(i)}(j)}_{\ell-i-2} = \floors{t^{(1)}}_{\ell-i-2}$ for all $j\in [n]$. Thus, in iteration $i+1$ we remain in state 1.\\ 
Otherwise, let $o \in [\ell-i-1]$ be the largest index such that $t_{\ell-i-1-o} \neq c-1$, and hence, $t^{(2)}_{\ell-i-1-k} = c-1$ for all $k\in \{0,\dots, o-1\}$. We have $\abs{\floors{t^{(1)}}_{\ell-i-1} + c^{-\ell+i+1}}_c = l-i-1-o$ and $\abs{\floors{t^{(1)}}_{\ell-i-1}}_c = l-i-1$ as well as $\floors{t^{(1)}}_{\ell-i-1-o-1} = \floors{t^{(2)}}_{\ell-i-1-o-1}$.  Thus, in iteration $i+1$ we are in state 2.

Assume now that after iteration $i$ the algorithm is in state 2 and let $t^{(1)}$ and $t^{(2)}$ be defined as above for state 2. Note that we have $q^{(i+1)} = t^{(1)}$ for all $j\in [n]$ with $q^{(i)} = t^{(1)}$. Hence, we get either
\begin{align*}
q^{(i+1)}(j) &= \floors{t^{(2)}}_{\ell-i-1}+ c^{-\ell+i+1}\\ 
		&= t^{(2)} - (c-1)c^{-\ell+i} + c^{-\ell+i+1} = t^{(1)} 
\end{align*}
 or $q^{(i+1)}(j) = \floors{t^{(2)}}_{\ell-i-1}$ for all $j \in [n]$. If $o = 1$, $\floors{t^{(1)}}_{\ell-i-2} = \floors{t^{(2)}}_{\ell-i-2}$. Hence, in iteration $i+1$ we are in state 1. If $o > 1$, we remain in state 2. 

Since in both possible states the number of different values of $q^{(i)}$ is at most 2 for all $i \in \{0, \dots, \ell\}$, the claim follows.
\end{proof}

We can now prove Theorem \ref{main lemma}.

\begin{proof}[Proof of Theorem \ref{main lemma}]

Remember that for $c$ even the theorem follows directly from Theorem \ref{weighted vs. normal} and Lemma \ref{simple case lemma}. Hence we assume that $c$ is odd for the rest of the proof. 

Let $z\in[0,1]$ have $c$--ary length $\ell \in \mathbb{N}$ and $p = z \mathbf{1}_n$. We iteratively round $q^{(0)} := p$ to colorings $q^{(i)}$ such that after each iteration $i \in [\ell]$, the $c$-ary length of $q^{(i)}(j)$ is at most $\ell - i$, while the rounding error $d_A(q^{(i-1)},q^{(i)})$ remains small. At the end of this process, we obtain a `pure' $c$--coloring $q:=q^{(\ell)}$.

The sequence $(q^{(i)})_{i\in [\ell]}$ is  defined in the obvious way. Remember $q^{(0)} = p \in C_{c,\ell}^{[n]}$. Having defined $q^{(i)} \in C_{c,\ell-i}^{[n]}$ for $i < \ell$, we apply Lemma \ref{one round in general} on $q^{(i)}$ to get $q^{(i+1)} \in C^{[n]}_{c,\ell-i-1}$ such that $d_A \left(q^{(i)},q^{(i+1)}\right) \leq  \frac{1}{2}c^{-\ell+i+2} \tau_{i} \herdisc\left(A, c\right)$ with $\tau_{i}$ being the number of different values of $q^{(i)}$. By Lemma \ref{number of different values is two}, we have $\tau_i \le 2$ for all $i \in [\ell]$.

We compute

\begin{align*}
d_{A}(p,q) & \leq  \sum_{i=0}^{\ell-1} d_A(q^{(i)},q^{(i+1)})\\
& \leq   \sum_{i=0}^{\ell-1} \tfrac{1}{2}(c-1)c^{-\ell+i+1} \tau_i \herdisc(A, c)\\
& \leq   c ~\mbox{herdisc}\left(A, c\right)
\end{align*}

Notice that we have restricted ourselves to $z \in [0,1]$ having finite $c$--ary length. Since $z \mapsto \herwdisc(A,z {\mathbf 1}_n)$ is continuous and the set of numbers having finite $c$--ary expansion is dense in $\left[0, 1\right]$, our bound carries over to arbitrary $z$.

\end{proof}

\section{A Tight Example}\label{seclower}

We now construct an infinite family of hypergraphs for which the bound of Theorem \ref{main lemma} is tight apart from constant factors.

Let  $c\in\mathbb{N}_{> 2}$, $k \in N_{\geq 2}$ and $n := 2ck$. We consider the complete hypergraph $\mathcal{H} = ([n],2^{[n]})$ on $n$ vertices.  Note that for complete hypergraphs, all induced subhypergraphs have a discrepancy of at most the discrepancy of the complete hypergraph. Hence, the hereditary discrepancy equals the discrepancy of $\mathcal{H}$.

As is easily seen, a $c$--coloring that partitions the vertex set in $c$ classes of equal sizes is optimal. The discrepancy is now witnessed, e.g., by monochromatic edges of size $n/c$, implying $\herdisc(\mathcal{H},c) = (1-\tfrac{1}{c}) n/c$. On the other hand, we have $\herdisc(\mathcal{H},2) = n/4$. Hence, we have $\herwdisc(\mathcal{H},2)\geq \herdisc(\mathcal{H}, 2) \geq \tfrac{1}{4}c \herdisc(\mathcal{H},c)$.

\section{Summary and Outlook}

In \cite{doerr-hereditary}, it was shown that the \mbox{hereditary} discrepancies of a hypergraph in different numbers of colors differ by at most  constant factors, which only depend on the number of colors involved. 

We improved this result by showing that we can bound the hereditary discrepancy in $b$ colors by $O(a)$ times the hereditary discrepancy in $a$ colors. Furthermore, such linear dependencies on the number of colors actually occur.

This work raises a number of open questions. A natural one is a determination of the absolute constant involved. Theorem~\ref{main lemma} shows the bound $\herwdisc(A,2) \le c \herdisc(A,c)$. In Section~\ref{seclower}, a class of hypergraphs fulfilling $\herwdisc(\HH,2) \ge \tfrac 14 c \herdisc(\HH,c)$ was presented. We currently do not know if the absolute constant of $1$ in Theorem~\ref{main lemma} can be improved towards $\frac 14$.

Possibly more interesting is the dependence on the second number of colors involved. Since we did a detour through $2$ colors, we only obtained the bound $\herdisc(A,b) = O(a) \herdisc(A,a)$. However, all examples we know only witness a bound of $\herdisc(A,b) = O(a/b) \herdisc(A,a)$. To show such a bound, it would be necessary to avoid the detour through $2$ colors, since the bound $\herdisc(A,b) = \Theta(1) \herdisc(A,2)$ cannot be improved. Avoiding the detour would also be of independent interest, as would a direct proof avoiding the rounding approach.

Another issue is the role of the linear discrepancy.  The $2$--color linear discrepancy of a matrix $A$ is \[\lindisc(A,2) := \max_{p : [n] \to [0,1]} \min_{q: [n] \to \{0,1\}} \max_{i \in [m]} \bigg|\sum_{j=1}^n a_{ij} (p(j) - q(j)) \bigg|.\] Hence the linear discrepancy problem is an extension of the weighted one where each vertex/column has an individual weight. In~\cite{doerr-hereditary}, $\lindisc(A,2) = O(c^2) \herdisc(A,c)$ was shown. It seems to be an interesting problem whether the linear discrepancy problem is harder than the weighted one in that we need this quadratic dependence or not. Again, we have no clue.


\renewcommand{\baselinestretch}{1}		

\small\normalsize						


\bibliographystyle{alpha}


\begin{thebibliography}{LSV86}

\bibitem[BF81]{bf}
J.~Beck and T.~Fiala.
\newblock ``{I}nteger making'' theorems.
\newblock {\em Discrete Applied Mathematics}, 3:1--8, 1981.

\bibitem[BS84]{BS}
J.~Beck and J.~Spencer.
\newblock Integral approximation sequences.
\newblock {\em Math. Programming}, 30:88--98, 1984.

\bibitem[BS95]{sos}
J.~Beck and V.~T. S\'os.
\newblock Discrepancy theory.
\newblock In R.~Graham, M.~Gr\"otschel, and L.~Lov\'asz, editors, {\em Handbook
  of {C}ombinatorics}, pages 1405--1446. Elsevier, 1995.

\bibitem[Doe02]{doerr-disc}
B.~Doerr.
\newblock Discrepancy in different numbers of colors.
\newblock {\em Discrete Math.}, 250:63--70, 2002.

\bibitem[Doe04]{doerr-hereditary}
B.~Doerr.
\newblock The hereditary discrepancy is nearly independent of the number of
  colors.
\newblock {\em Proc.\ Amer.\ Math.\ Soc.}, 132:1905--1912, 2004.

\bibitem[DS03]{wirmcol}
B.~Doerr and A.~Srivastav.
\newblock Multicolour discrepancies.
\newblock {\em Combinatorics, Probability and Computing}, 12:365--399, 2003.

\bibitem[Gho62]{ghou}
A.~Ghouila{-H}ouri.
\newblock Caract\'erisation des matrices totalement unimodulaires.
\newblock {\em C.~R.~Acad. Sci. Paris}, 254:1192--1194, 1962.

\bibitem[LSV86]{LSV}
L.~Lov\'asz, J.~Spencer, and K.~Vesztergombi.
\newblock Discrepancy of set-systems and matrices.
\newblock {\em Europ. J.~Combin.}, 7:151--160, 1986.

\bibitem[Spe85]{spencer}
J.~Spencer.
\newblock Six standard deviations suffice.
\newblock {\em Trans. Amer. Math. Soc.}, 289:679--706, 1985.

\end{thebibliography}

\end{document}